\documentclass[useAMS,usenatbib]{mn2e}

\title[Applying CREAM to AGN Light Curves]{Accretion Disc Time Lag Distributions: Applying CREAM to Simulated AGN Light Curves}
\author[D. A. Starkey et al.]{{D. A. Starkey$^{1}$\thanks{E-mail:
ds207@st-andrews.ac.uk (DAS); kdh1@st-andrews.ac.uk (KDH); cv21@st-andrews.ac.uk (CV)}, Keith Horne$^{1}$, C. Villforth$^{1,2}$}\\
$^{1}$SUPA School of Physics and Astronomy, St Andrews, KY16 9SS, Scotland, UK\\
$^{2}$University of Bath, Department of Physics, Claverton Down, Bath, BA2 7AY, UK\\
}

\usepackage{booktabs}
\usepackage{anysize}
\usepackage{graphicx}
\usepackage{epsfig}
\usepackage{float}
\usepackage{natbib}
\usepackage{setspace}
\usepackage{subfigure}
\usepackage{caption}
\usepackage[section]{placeins}
\usepackage{rotating}
\usepackage{placeins}
\usepackage[margin=0.5in]{geometry}
\usepackage[table,xcdraw]{xcolor}
\usepackage{amsmath}
\usepackage{url}
\usepackage{gensymb}
\usepackage{color}

\newcommand{\mmdot}{{M\dot{M}}}

\begin{document}

\date{Accepted 2015 November 18. Received 2015 November 12; in original form 2015 September 17}

\pagerange{\pageref{firstpage}--\pageref{lastpage}} \pubyear{2002}

\maketitle

\label{firstpage}

\begin{abstract}
Active Galactic Nuclei (AGN) vary in their brightness across all wavelengths. Moreover, longer wavelength ultraviolet - optical continuum light curves appear to be delayed with respect to shorter wavelength light curves. A simple way to model these delays is by assuming thermal reprocessing of a variable point source (a lamp post) by a blackbody accretion disc. We introduce a new method, CREAM (\textbf{C}ontinuum \textbf{RE}processed \textbf{A}GN \textbf{M}arkov Chain Monte Carlo), that models continuum variations using this lamp post model. The disc light curves lag the lamp post emission with a time delay distribution sensitive to the disc temperature-radius profile and inclination. We test CREAM's ability to recover both inclination and product of black hole mass and accretion rate $\mmdot$, and show that the code is also able to infer the shape of the driving light curve. CREAM is applied to synthetic light curves expected from 1000 second exposures of a 17th magnitude AGN with a 2m telescope in Sloan g and i bands with signal to noise of 500 - 900 depending on the filter and lunar phase. We also tests CREAM on poorer quality g and i light curves with SNR = 100. We find in the high SNR case that CREAM can recover the accretion disc inclination to within an uncertainty of 5 degrees and an $\mmdot$ to within 0.04 dex.
\end{abstract}


\section{Introduction}

Active Galactic Nuclei (AGN) at the centres of distant galaxies are thought to contain an accretion disc orbiting a supermassive black hole (SMBH), clouds of both broad and narrow-line-emitting gas (the BLR and NLR) that surround the AGN with uncertain geometry and kinematics, and a dusty torus that obscures the disc component when viewed close to edge on \citep{ur95,el00}. The accretion disc emission is thought to give rise to the ``Big Blue Bump'' feature observed in the spectra of AGN and quasars \citep{ma83,sh05}. The BLR extents over 10 - 200 light days \citep{be13} and we can infer from observed continuum - continuum light curve time delays \citep{ed15,ca07} and light-travel-time arguments that the accretion disc occupies a region several light days in size. AGN distances make resolving their structure impractical at this time. A number of studies \citep{ho14,ki11} use infra-red interferometry to resolve the inner edge of the dusty torus, but all AGN are too remote to resolve the BLR much less the accretion disc. To obtain information on the accretion disc, one can therefore proceed along one of two routes. The half light radii of the accretion disc at some wavelength can be inferred from microlensing observations of gravitational lensed quasars \citep{ma15,mo12}. Lensing produces multiple images of the quasar whose flux ratios are sensitive to the size of the emitting region at the specified wavelength. One can alternatively obtain information on the structure and kinematics of the BLR and accretion disc in the time rather than space domain via Reverberation Mapping (RM) \citep{bl82}. RM assumes that changes in the accretion disc emission drive the BLR line variability and, by measuring time delays between the disc and broad line emission, one can infer the black hole mass and size of the BLR.

As well as continuum - line delays, one can perform RM studies between continuum light curves at different wavelengths and this is the subject of this paper. AGN and quasar continuum emission exhibits time variability that can be modelled by a damped random walk (DRW) \citep{ke09}. A DRW is characterised by a stochastic differential equation whose power spectrum is a broken power law with a slope that steepens from 0 to -2 above a break timescale. In AGN observed in the sloan digital sky survey (SDSS), the break timescale is upwards of 100 days, correlates positively with back hole mass and wavelength and is a weak function of luminosity \citep{ma10}.

Light curves at longer wavelengths appear to lag behind those at shorter wavelengths \citep{ca07,ch13,ed15} with evidence of a mean delay scaling with wavelength like $\langle \tau \rangle \propto \lambda^{4/3}$. The inhomogeneous disc model suggests that variability arises from temperature fluctuations that occur stochastically throughout the accretion disc \citep{de11}. While this can explain the magnitude and timescales of variability observed by e.g \citet{ma10}, it does not easily explain the short-to-long-wavelength time lag. This feature emerges naturally if the disc is irradiated by hard photons close to the black hole, with time delays increasing toward longer wavelengths due to light travel time effects. It is not known whether the photons driving the variability are emitted at far UV or harder X-ray wavelengths. SWIFT observations of NGC 5548 \citep{mc14} show evidence of X-rays leading UV but accurate models of the delay distribution function are crucial for understanding the origin of disc variability.

Cross correlating short and long wavelength light curves is the traditional way to measure both continuum - line and continuum - continuum time delays. Two cross correlation techniques to measure lags despite the uneven sampling (for example incurred from bad weather, blockage by the sun or observing time constraints) are the interpolated cross correlation function (ICCF) \citep{ga87} and discrete cross correlation function (DCCF) \citep{ed88}. ICCF works by linearly interpolating uneven data into evenly spaced data, whereas the DCF bins the uneven data into equal time widths. The peak or centroid of the CCF gives an estimate of the mean time delay.

A more recent development, SPEAR \citep{zu11} (and the python implementation JAVELIN) fits the DRW model to an input light curve that is assumed to drive the AGN variability. The driver is usually assumed to be the continuum light curve observed at the smallest wavelength. SPEAR then smooths and delays this light curve in an attempt to best model the continuum and line light curves. This method has been successfully applied to model both continuum - line delays \citep{zu13} and continuum - continuum delays \citep{ed15}. Time delays measured from CCF and SPEAR are in general agreement with one another but each of these methods returns just a single number for the time delay (e.g the mean, CCF peak, or centroid of an assumed top hat delay function).

In this work, we present a Markov Chain Monte Carlo (MCMC) approach to fit a reverberating accretion disc model and test this using synthetic light curves. Our approach differs from SPEAR and CCF routines as it returns a distribution of continuum time delays that is a function of accretion disc parameters such as the inclination and product $\mmdot$ of black hole mass $M$ and accretion rate $\dot{M}$. Furthermore, the MCMC code (CREAM) does not require an input light curve to act as the driver of variability, it instead infers the shape of the driving light curve.

The layout of the paper is as follows. Section \ref{sec2} describes the assumed accretion disc model. In Section \ref{secmethod} we detail the statistical framework of the MCMC code. We describe how the synthetic data sets used to test the code are constructed in Section \ref{secsynthdat}. Section \ref{secresults} presents results of applying CREAM to the synthetic light curves and shows the posterior probability distributions for $\mmdot$ and disc inclination. We present discussion and conclusions in Section \ref{secdiscussion}.

\section{Accretion Disc Thermal Reprocessing Model}
\label{sec2}

We model continuum light curve variability as temperature fluctuations due to variable irradiation of a flat, blackbody accretion disc. Variability arises due to accretion disc reprocessing of photons from a variable point source located just above the black hole. In this section, we describe the underlying physics of the model and introduce the time delay distribution appropriate to our accretion disc model.

\subsection{Steady State Temperature Radius Structure}
\label{secbackgrounddisk}

\begin{figure}
\includegraphics[scale=0.32]{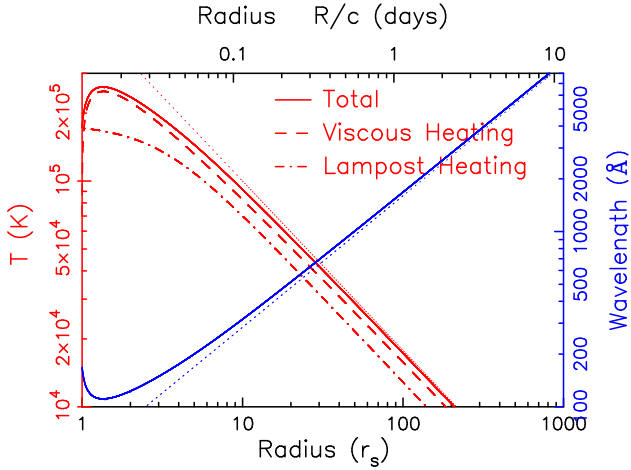}
\caption{Temperature radius profile due to the viscous heating (dashed red line) and lamp post heating (dot-dashed red line). The total effect is shown in solid red and the dotted red line shows a power law slope of $-3/4$. The red and blue lines are described by the y-axis on the left and right hand sides of the plot respectively. The solid blue line shows the peak wavelength in the blackbody spectrum $B_\nu \left( \lambda_\mathrm{peak} \right)$ as a function of radius assuming a temperature radius law given by Equation \ref{eqtrprof}. For a flat disc $\lambda_\mathrm{peak} \propto r^{4/3}$. We plot a 4/3 line in dotted blue for comparison. The plot is appropriate for a SMBH of $10^8M_{\odot}$ accreting at $1 M_{\odot} \mathrm{yr}^{-1}$.}
\centering
\label{figtrprof}
\end{figure}

The thermal reprocessing model assumes that each point in a flat, optically thick accretion disc has an associated blackbody temperature $T$ and emits at wavelength $\lambda$ according to the Planck function,

\begin{equation}
\label{eq_planck}
 B_\nu \left( \lambda , T \right) = \frac{2 h c}{\lambda^3}\frac{1}{e^{hc/ \lambda k T}-1},
 \end{equation}

\noindent where $h$ and $k$ are the Planck and Boltzmann constants, and $c$ is the speed of light. We model the source of irradiation as a ``lamp post'' with bolometric luminosity $L_b (t)$, located a height $h_x$ above the black hole. The lamp post approximates the region driving the variability as a point source from which disc variability at radius $r$ and azimuth $\phi$ is delayed by a single light-travel-time induced lag. To compute the disc flux at a given wavelength, we require a temperature-radius relation. The $T(r)$ profile (Figure \ref{figtrprof}) combines the effects of viscous heating due to differential rotation and irradiation by the lamp post and can be expressed as \citep{AP,ca07}

\begin{equation}
T^4=\frac{3GM\dot{M}}{8\pi \sigma r^3}\left(1-\sqrt{\frac{r_{in}}{r}}\right) + \frac{L_{b}(1-a)h_x}{4{\pi}{\sigma}x^3}, 
\label{eqtrprof}
\end{equation}

\noindent where $M$ is the black hole mass, $\dot{M}$ the accretion rate, $a$ the disc albedo, $\it{r}$ the orbital radius of the surface element, $h_x$ the lamp post height, $x = \sqrt{r^2 + h_x^2}$ the distance from the lamp post point to surface element and $r_{\mathrm{in}}$ is the inner most stable circular orbit (3 $r_s$ for a Schwarzschild black hole). In the limit $r >> r_{\mathrm{in}} , h_x$, Equation \ref{eqtrprof} simplifies to

\begin{equation}
\label{eqtrprofparm}
T= T_{0} \left( \frac{r_0}{r} \right)^{3/4},
\end{equation} 

\noindent where

\begin{equation}
T_{0}^4 = \frac{3GM\dot{M}}{8\pi \sigma r_0^3} + \frac{h_x \left( 1 - a \right) L_b}{4 \pi \sigma r_0^3}.
\end{equation}

\noindent The lamp post's bolometric luminosity is $L_b = \eta \dot{M} c^2$ with an efficiency parameter $\eta$  typically around 0.1 \citep{sh09}.

The X-ray emitting region has been constrained by microlensing techniques to within just 6 $r_s$ of the SMBH \citep{mo12} and \citet{mc15} find evidence of X-ray variability leading that at UV wavelengths. The point source lamp post approximation can therefore be justified with reference to Figure \ref{figtrprof}. The shortest wavelength light curve we use in this work is in the sloan u filter with central wavelength 3545 $\AA$. For a fiducial case black hole with $M = 10^8 M_\odot$ and $\dot{M} = 1 M_\odot \mathrm{yr}^{-1}$ the radius of peak emission at this wavelength is $\sim 300 r_s$ (Figure \ref{figtrprof}). The 6 $r_s$ X-ray emitting region effectively appears as a point source at this radius.

\begin{figure}
\center
\includegraphics[scale=0.32]{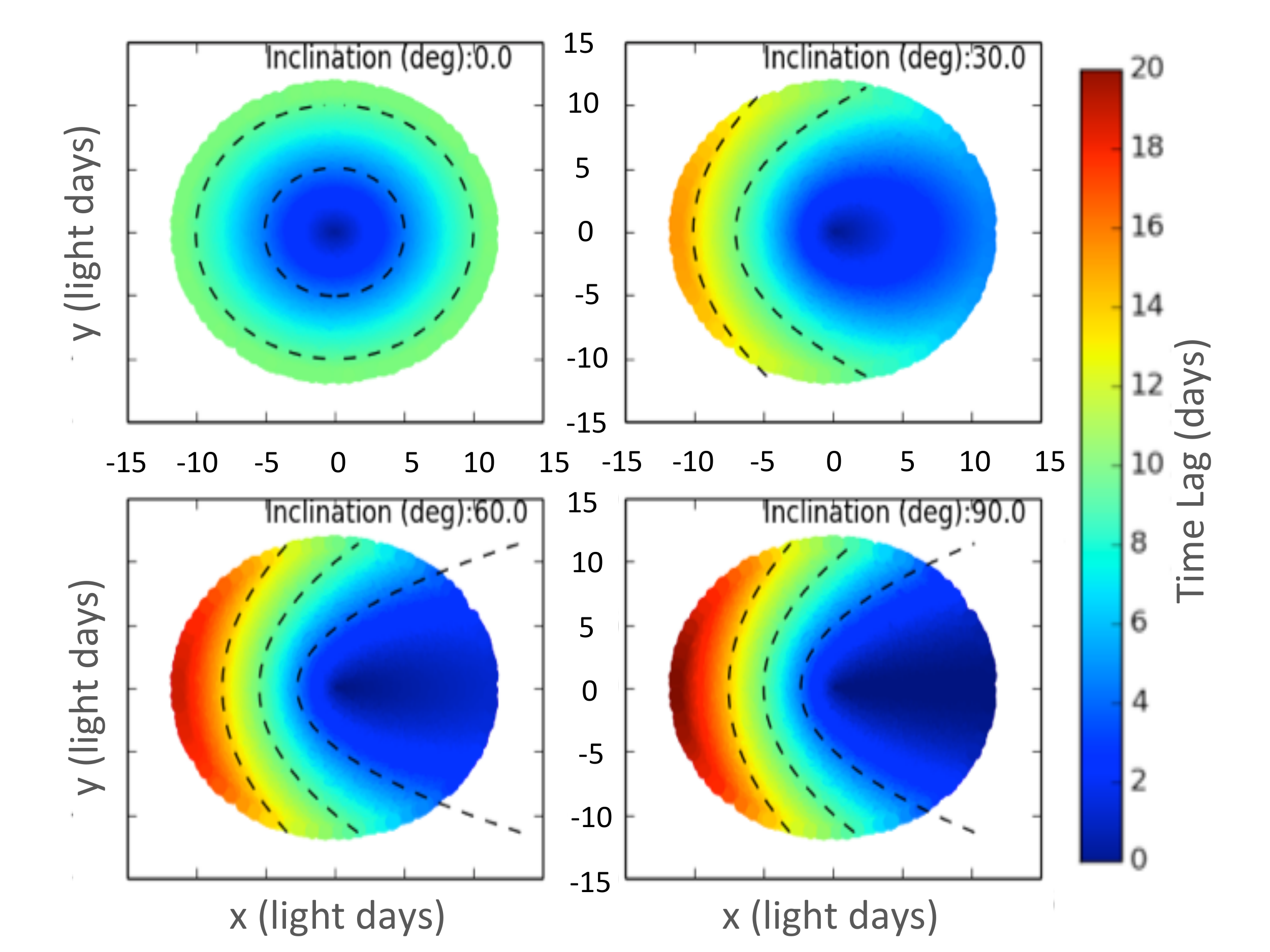}
\caption{Accretion disc time delay profiles $\tau \left(r, \phi, i \right)$ evaluated at $0^o$ (top left), 30$^o$ (top right), $60^o$ (bottom left), $90^o$ (bottom right). Dashed lines represent 5, 10 and 15 day isodelay surfaces.}
\label{figtdel}
\end{figure}

\begin{figure}
\center
\includegraphics[scale=0.28,angle=0,trim=0cm 0cm 0cm 0cm]{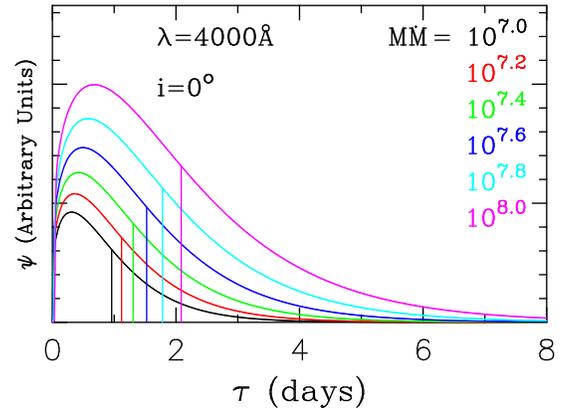}
\includegraphics[scale=0.28,angle=0,trim=0cm 0cm 0cm 0cm]{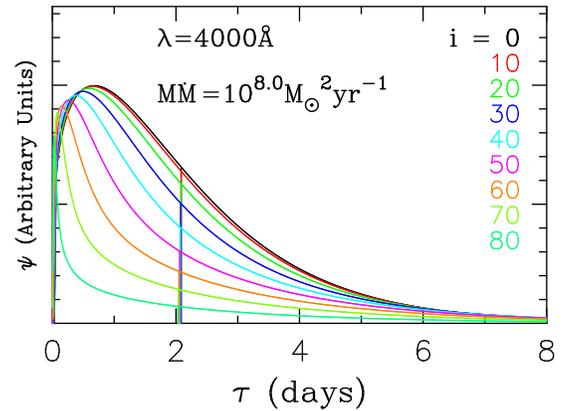}
\includegraphics[scale=0.28,angle=0,trim=0cm 0cm 0cm 0cm]{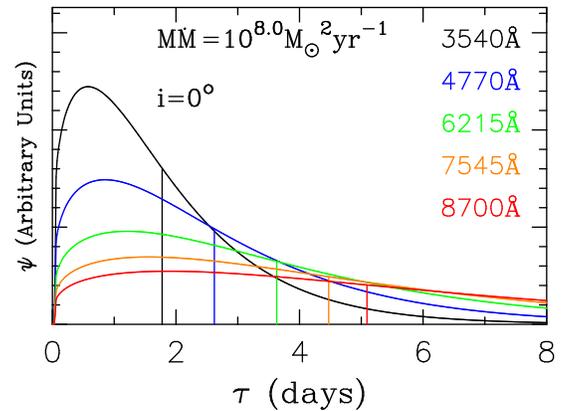}
\caption{The numerically evaluated response functions $\psi \left( \tau | \lambda \right)$ for a face-on, flat disc around a $10^8M_{\odot}$ black hole accreting at $1 M_\odot \mathrm{yr}^{-1}$. The lamp post is set at $h_x = 3r_s$ above the disc plane. We then show the effect of varying $M\dot{M}$ (top), inclination (middle) and wavelength $\lambda$ (bottom). Vertical lines show the mean delay $\langle \tau \rangle$.}
\label{figtf}
\end{figure}

\subsection{Disc Time Delays}

\label{sectimevariability}
Photons emitted isotropically by the lamp post intercept the disc surface. Once absorbed, the lamp post photons are assumed to heat the disc instantaneously. This is known as an instantaneous thermal reprocessing model \citep{ca07} and causes the disc at the location of impact to radiate as a blackbody with a higher temperature. The reprocessed photons are seen by a distant observer with a light-travel-time-induced lag, $\tau \left( r, \phi, i \right)$, dependent on the radius, $r$, and azimuth, $\phi$, of reprocessing site and inclination, $i$, of the disc with respect to the observer ($i=0$ corresponds to a face on disc). For a flat disc, 

\begin{equation}
\label{eq_tauflat}
\tau \left( r, \phi \right) = \frac{r}{c} \left( 1 + \cos \phi  \sin i \right).
\end{equation}

\noindent Surfaces of constant time delay form circles for a face on disc ($i=0^o$), parabolas for an edge on disc and ellipses for an intermediately inclined disc (Figure \ref{figtdel}).

\subsection{Disc Response Function}
Figure \ref{figtdel} shows that the continuum reprocessing occurs at reprocessing sites over a range of radii $r$ and azimuths $\phi$ corresponding to a distribution of time lags. To obtain the total disc flux $F \left( \lambda, t \right)$ we convolve the variable component of the driving light curve flux $\Delta F_x \left( t \right)$ with a time delay distribution $\psi\left( \tau \vert \lambda \right)$ and add on a background spectrum $\overline{F_\nu} \left( \lambda \right)$,

\begin{equation}
\label{eqecho}
F_{\nu}(\lambda , t)= \overline{F}_{\nu} \left( \lambda \right)  + \Delta F_{\nu}(\lambda)  \sum_{i=0}^{\tau_\mathrm{max}} \psi \left( \tau_i | \lambda \right) \Delta F_{x}(t- \tau_i ) \Delta \tau.
\end{equation}

\noindent Here ${\psi} \left( {\tau \vert \lambda}\right)$ is dimensionless and normalised such that $\sum {\psi} \left( {\tau \vert \lambda}\right) \delta \tau = 1$ and $\Delta F_{\nu} \left( \lambda \right)$ is the variable component of the spectrum. The response function $\psi \left( \tau | \lambda \right)$ for a flat disc is derived in \citet{ca07} and is a function of $\mmdot$ and inclination. We plot several example response functions showing the $\mmdot$, inclination and wavelength dependences in Figure \ref{figtf}. The mean delay, as shown by the vertical lines in Figure \ref{figtf}, is independent of inclination. Tilting the disc makes the response function more skewed; it peaks at shorter lags and develops a tail toward large lags. This effect arises because higher inclinations increase the lag on the far side and decrease the lag of the near side of the disc relative to a face on inclination (see Figure \ref{figtdel}). If we increase $\mmdot$, reprocessing sites emitting at a fixed temperature move to larger radii. This scales the response function without altering its shape with mean delays scaling like $\left< \tau \right> \propto \left( M\dot{M} \right) ^{1/3} \lambda^{4/3}$. Both the mean delay and response function width increase with wavelength. These effects occur due to the reduction in temperature with radius (Equation \ref{eqtrprofparm}). We see from Figure \ref{figtrprof} that the assumed blackbody-emitting disc allows the longer wavelength emission to extend to larger radii with higher time delays.

\section{CREAM: Continuum Reprocessing Echo AGN Markov chain monte carlo code}
\label{secmethod}

With the effects of disc inclination and black hole $M\dot{M}$ in mind, we fit a multi-parameter model to a set of light curves. Our Continuum Reprocessing Echo AGN Markov Chain Monte Carlo code (CREAM) is detailed in these subsections. CREAM ingests light curve data at $N_\lambda$ wavelengths and uses a Metropolis-within-Gibbs parameter sampling approach to sample the posterior parameter distributions and find the best fitting parameter values and uncertainties (for both $M\dot{M}$, inclination and the parameters in the driving light curve). CREAM fits Equation \ref{eqecho} to an input set of continuum light curves. The mean spectrum $\overline{F}_{\nu}(\lambda)$ models all constant contributions to these echo light curves. This primarily addresses host galaxy contamination but considers also any other slowly varying or invariant emission sources. The $\Delta F_{\nu} \left( \lambda \right)$ parameters scale the echo light curve variations to match the echo light curve data.

\subsection{Driving Lightcurve: $\Delta F_x \left( t \right)$}

Many CCF studies of continuum light curves assume the shortest wavelength continuum light curve acts as the driver of variability. Lags of optical to X-ray light curves have been observed in several \citep{ar08,mc15} objects but in other instances X-rays appear only weakly to relate to UV and optical variability \citep{ed15}. It is therefore not fully understood what wavelength regime acts as the variability-driving lamppost. CREAM has the ability to infer the shape of the true driving light curve from the shape of the echo light curves alone. In order to achieve such a realisation of the driving light curve, we need parameters that fully specify the shape of the driving light curve and an ability to constrain these parameters so that the observed power spectrum is consistent with observations. We represent the driving light curve by a Fourier time series, 

\begin{equation}
\label{eqdrive}
\Delta F_{x} \equiv F_{x}\left( t \right) - \overline{F}_x  = \sum\limits_{k=1}^{N_k} C_k \cos(\omega_k t) + S_k \sin(\omega_k t),
\end{equation}

\noindent where $\omega_k = k \Delta \omega$ is the $k$th Fourier frequency, $S_k$ and  $C_k$ are the Fourier amplitude coefficients. $\overline{F}_x$ is the mean of the driving light curve. The reference level $\overline{F_x}$ is therefore somewhat arbitrary and we use $\overline{F_x} = 0$ for our tests. If however we choose to include driving light curve data, we set the Fourier sum in Equation \ref{eqdrive} to correspond to the $\log$ of the driving light curve variable component. This enforces positivity in $\Delta F_{x}\left( t \right)$ that then oscillates about 1. The reference level $\overline{F}_x$ is then included as a multiplicative parameter. The low and high frequencies ($\omega_{\mathrm{low}}$ and $\omega_{\mathrm{hi}}$) in the Fourier sum (Equation \ref{eqdrive}) can be specified as the code is initialised. For these tests we use,

\begin{equation}
\label{eqomegalo}
\omega_{\mathrm{low}} = \Delta \omega = \frac{1}{2} \frac{2\pi}{T_\mathrm{rec}},
\end{equation}
\begin{equation}
\label{eqomegahi}
\omega_{\mathrm{hi}} = \frac{2 \pi}{\overline{\Delta t}} = N_k \Delta \omega.
\end{equation}

\noindent Here $T_\mathrm{rec}$, the recurrence time of the Fourier series, should be longer than the timespan of the data plus the width of the response function (Figure \ref{figtf}). $\overline{\Delta t}$ is the mean time separation between adjacent points. For the specific tests in this work, our light curves span 100 days and we choose $T_\mathrm{rec} = 200$ days. We discuss the choice of upper frequency in Section \ref{secdiscussion} but the default value specified in Equation \ref{eqomegahi} satisfies the Nyquist sampling requirement of fewer than 2 data points within 1 period of the high frequency component.

\subsection{The Badness of Fit (BOF)}

Adopting a Bayesian framework, we begin this discussion with Bayes' theorem, relating the posterior probability distribution, $P( \mathbf{\Theta} | \mathbf{D} )$, of $N_\mathrm{par}$ parameters, $\mathbf{\Theta} = \left( \theta_1 ... \theta_{N_\mathrm{par}} \right)$, with priors $P( \mathbf{\Theta} )$, fitted to data, $\mathbf{D} = \left( D_1 ... D_N \right)$,

\begin{equation}
\label{eqbayes}
P(\mathbf{\Theta} | \mathbf{D}) = P \left( \mathbf{D} | \mathbf{\Theta} \right) \frac{P \left( \mathbf{\Theta} \right) }{P \left( \mathbf{D} \right) }.
\end{equation}

\noindent Here $P(\mathbf{D}| \mathbf{\Theta} )$ is the `likelihood function', $P(\mathbf{D})$ normalises the posterior probability

\begin{equation}
\label{eqbay_norm}
\int P( \mathbf{\Theta} |\mathbf{D}) \mathrm{d} \mathbf{\Theta} = 1
\end{equation}

\noindent The `Badness of Fit' (BOF) is defined as $-2 \, \mathrm{ln} \left( P(\mathbf{\Theta} |\mathbf{D}) \right)$ and takes the form

\begin{equation}
\label{eqBOF_unpri}
\mathrm{BOF} = \chi^2 + \sum_{i=1}^N \ln \left( \sigma_i^2 \right) -2 \ln \left( P \left( \mathbf{\Theta} \right) \right) + \mathrm{const},
\end{equation}

\noindent where 
\begin{equation}
\chi^2=\sum\limits_{i=1}^{N} \left( \frac{D_i-M_i}{\sigma_i} \right)^2.
\label{eqchisq}
\end{equation}

\noindent The model $\mathbf{M} = (M_1 ... M_N )$ is a function of the parameters $\mathbf{\Theta}$, and we assume Gaussian errors on the $N$ light curve data points with errorbars $\sigma_i$.

\subsection{Priors $P\left( \mathbf{\Theta} \right)$}
\label{secpriors}

The model parameters and their priors are summarised in Table \ref{tabprior}. Counting the model parameters, we have $2N_k$ sine and cosine amplitudes specifying the shape of the driving light curve about a mean level $\overline{F}_x$. We also have a response function parametrised by $i$ and $\mmdot$ that smooths and delays the driving light curve. The mean flux $\overline{F}_{\nu}\left( \lambda \right)$ and scaling terms $\Delta F_{\nu} \left( \lambda \right)$ map the smoothed-delayed driving light curve onto the reprocessed-echo light curve data. In total, $N_\mathrm{par} = 2 \left( N_k + N_\lambda + 1 \right) + 1$.

The inclination prior is uniform in $\cos i$ reflecting the assumption that the disc is randomly orientated in the sky. The Fourier amplitudes, $S_k$ and $C_k$, are constrained with Gaussian priors as described in Equations \ref{eqpriorsimp} to \ref{eqbof}. For necessarily positive parameters $\mmdot$, $\Delta F_{\nu} \left( \lambda \right)$ and $\overline{F}_{\nu}\left( \lambda \right)$ our priors are uniform in $\log$. 

\begin{figure*}
\includegraphics[scale=0.87,angle=0,trim=5cm 3.5cm 5.8cm 2.5cm]{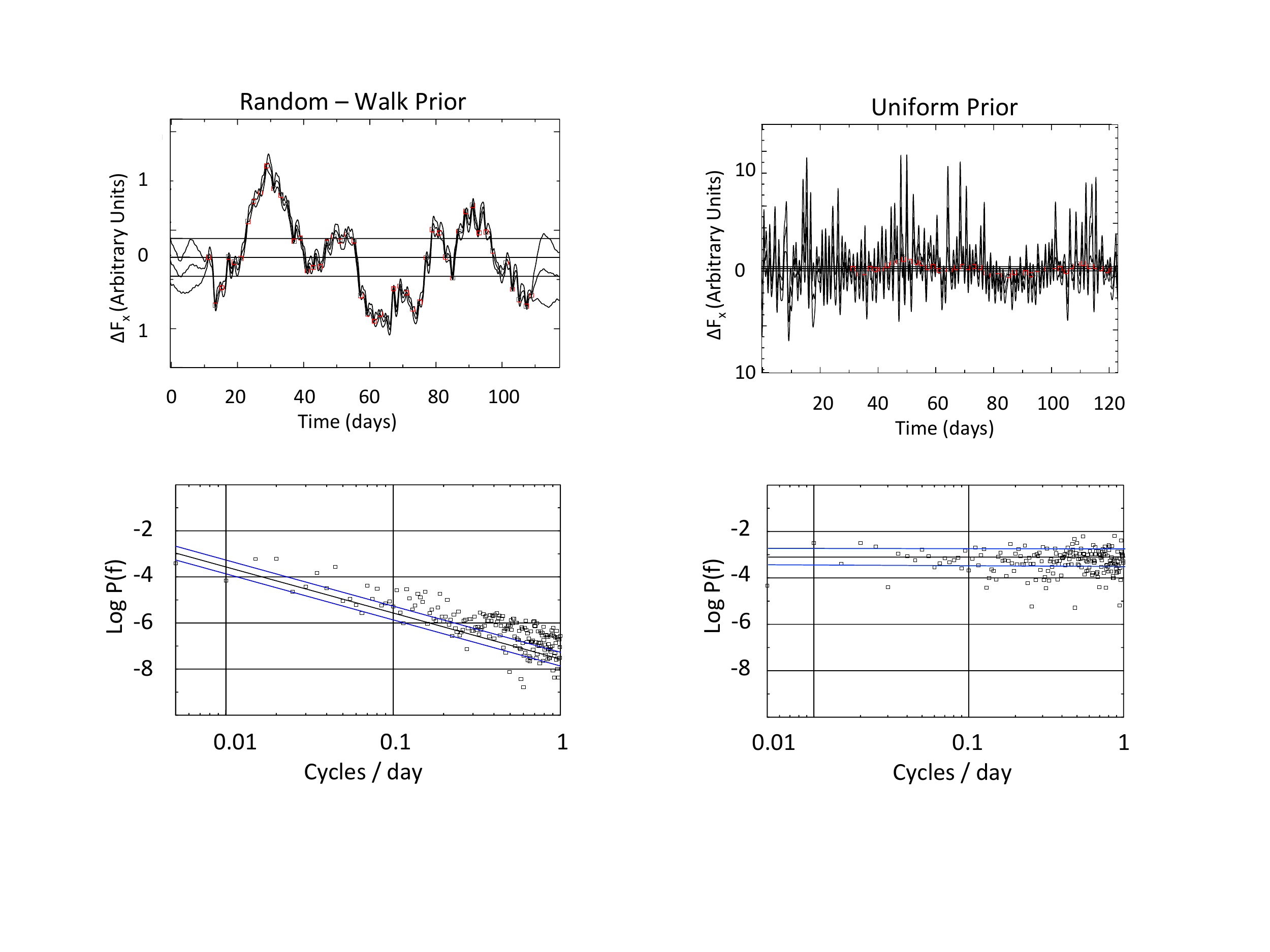}
\caption{Left: CREAM fit to synthetic light curve with data points in red and mean $\pm$ 1 $\sigma$ envelopes in black. Horizontal lines indicate the reference position with $1 \sigma$ envelopes. The bottom panel shows the power spectrum of the resulting fit with the black squares showing $S_k^2 + C_k^2$ for each Fourier frequency. The black and blue lines indicate the random walk prior (Equation \ref{eqpriorsimp}). On the right we show the effect of fitting the same data with a uniform prior.}
\label{figcompare}
\end{figure*}

We now discuss our choice of the prior used to constrain the driving light curve. With uniform priors on the Fourier amplitudes the driving light curve over-fits the data by over utilizing the high frequency Fourier amplitudes. This effect is demonstrated in Figure \ref{figcompare} and motivates the choice of a more suitable prior to constrain the shape of the driving light curve. \citet{mc14} suggest that X-ray light curves may drive AGN variability and we choose to constrain the $S_k$ and $C_k$ parameters using a `random walk prior'. This prior drives the driving light curve's $S_k$ and $C_k$ Fourier parameters (Equation \ref{eqdrive}) towards values giving rise to a random walk power spectrum,

\begin{equation}
\label{eqpriorsimp}
\langle S_k^2 \rangle + \langle C_k^2 \rangle = P(\omega_k) \Delta \omega=P_0 \Delta \omega \left( \frac{ \omega_0 }{\omega_k} \right) ^{\alpha},
\end{equation}

\noindent with $\alpha = 2$ motivated by studies of X-ray variability \citep{ut02,mc06}.  We incorporate the power spectrum prior in Equation \ref{eqpriorsimp} as a Gaussian prior on the $S_k$ and $C_K$ parameters with mean $\langle S_k \rangle$ = $\langle C_k \rangle$ = 0, and variance $\langle S_k^2 \rangle = \langle C_k^2 \rangle = \sigma_k^2$,

\begin{equation}
\label{eqpriorfinal}
P \left( \bf{\Theta}\right) = \prod_{k =1}^{N_k} \frac{e^{- \frac{1}{2}\frac{C_k^2 + S_k^2}{{\sigma_k}^2}}}{2\pi \sigma_k^2},
\end{equation}

\noindent where,
\begin{equation}
\label{eqsigk}
\sigma_k^2 =\frac{P_0 \Delta \omega}{2}  \left( \frac{ \omega_0 }{\omega_k} \right) ^{2}.
\end{equation}

\noindent The $\mathrm{BOF}$, substituting Equation \ref{eqpriorfinal} in \ref{eqBOF_unpri}, is then

\begin{equation}
\label{eqbof}
\begin{split}
\text{BOF} & = \chi^2 + \sum_{i=1}^{N} \ln (\sigma_i^2)\\
& + \sum_{k=1}^{N_k} \left( 2 \ln (\sigma_k^2) + \frac{C_k^2+S_k^2}{\sigma_k^2} \right) + \mathrm{const}.
\end{split}
\end{equation}

\noindent This `random walk prior' gives a more satisfactory fit, as shown in Figure \ref{figcompare}.

\begin{table}
\center
\caption{Summary of priors on the CREAM parameters ($\sigma_k$ is defined in Equation \ref{eqsigk}). The table shows (left column) the parameter, (middle column) the number of relevant parameters and (right column) the prior on the parameter(s).}
\begin{tabular}{ccc}
\hline
Parameter & $N_\mathrm{par}$ & Prior\\
\hline 
$S_k$ and $C_k$ & 2$N_k$ & Gaussian ($\langle S_k \rangle = \langle C_k \rangle = 0$,  \\
&& $\langle S_k^2 \rangle = \langle C_k^2 \rangle = \sigma_k^2$)   \\
$\cos i$ & 1 & Uniform \\
$\log \mmdot$ & 1 & Uniform \\
$\log \Delta F_{\nu}$ & $N_{\lambda}$ & Uniform \\
$\log \overline{F}_{\nu}\left( \lambda \right)$ & $N_{\lambda}$ & Uniform \\
\hline
\end{tabular}
\label{tabprior}
\end{table}

\section{Synthetic Light Curves}
\label{secsynthdat}

Reverberation mapping observing campaigns require high cadence, long duration light curves. We choose to test CREAM on synthetic light curves (Figure \ref{figfake}) generated assuming the observing capabilities of the Las Cumbres Observatory Global Telescope (LCOGT) network in Sloan g and i filters with 1000s exposures. The global longitude coverage of the LCOGT network allows targets to be observed without daily obstruction by the Earth and the g and i filters offer both high signal to noise and broad wavelength coverage (4775$\AA$ - 7540$\AA$).

We first generate a uniform time grid $t_i = i\Delta t$ on which to evaluate the model light curves. This is convenient for evaluating the convolution integral (Equation \ref{eqecho}). We use $\Delta t =0.1$ days and generate the driving light curve using a random walk about a mean $\overline{F}_x = 0$, 

\begin{equation}
\label{eqRW}
\Delta F_x \left( t_i \right) \equiv F_x \left( t_i \right) - \overline{F}_x = \Delta F_x (t_{i-1}) + \sum_{k=1}^{i} G(0, \sigma \sqrt{\Delta  t}).
\end{equation}

\noindent Here $G(a,b)$ represents a random sample from a Gaussian distribution with mean $a$ and standard deviation $b$. We adopt $\sigma = 1$ for the standard deviation of the random walk after 1 day.  

The echo light curves are then obtained from the driver by convolution with the response function (Figure \ref{figtf}) evaluated at inclination $30^\circ$ and $\mmdot =10^8 M_{\odot}^2 \mathrm{yr}^{-1}$. These are the true parameters for our simulated light curves. The mean level of the g light curve is set at AB magnitude 17 and the mean level of the i light curve scales relative to this reference level like $F_\nu( \lambda ) \propto \lambda^{-1/3}$, as expected for a flat blackbody-emitting accretion disc. This brightness is appropriate for a target observed at $z = 0.27, \Omega_\Lambda = 0.7, \Omega_m = 0.3, H_0 = 70 \, \mathrm{km \, s}^{-1}\mathrm{Mpc}^{-1}$, with an accretion rate, $\dot{M} = 1M_\odot \mathrm{yr}^{-1}$ and a black hole mass $M = 10^8 M_\odot$, fairly typical for an AGN. 

\citet{ma10} models AGN continuum light curves as damped random walks with a characteristic break time scale $\tau_B$. We use the empirical relation between light curve variance, $z, \dot{M}, M$ and $\lambda$ (Equation 7 of \citet{ma10}) to determine the appropriate variability amplitude for our light curves. This equation returns a parameter $\mathrm{SF}_{\infty}$ that is related to the light curve variance by

\begin{equation}
\label{eqsfinf}
\sigma^2 \left( F_\nu \right) = \frac{\mathrm{SF}_\infty^2 }{2} \left( 1 - e^{t_\mathrm{len} / \tau_B} \right),
\end{equation}
where $t_\mathrm{len}$ is the length of the light curve. $\tau_B$ is again empirically constrained by \citet{ma10} in terms of the parameters $\lambda, z, \dot{M}$ and $M$ and we quote in Table \ref{tabsynthprop} the values of $\tau_B$ and $\mathrm{SF}_\infty$ used to construct our synthetic light curves.

\subsection{Observational Uncertainties}
We mimic irregular sampling over the 100 day observing period by linearly interpolating the 0.1 day cadence echo light curves onto a new light curve sampled at times $t_{i=1...N}$ according to

\begin{equation}
\label{eqransamp}
t_i = t_{i-1} + G\left(0,\overline{\Delta t} \right),
\end{equation}

\noindent where $\overline{\Delta t}$ is the mean time separation between adjacent points (for our light curves, we use $\overline{\Delta t} = 1$). We also require $t_i > t_{i-1}$.  

The error bars, generated using the LCOGT exposure time calculator \footnote{\url{http://lcogt.net/files/etc/exposure_time_calculator.html}}, are appropriate for observations of a 17th magnitude target observed in ugriz at airmass 1.3 with 1000 second exposures. We compare this with a first order signal to noise calculation assuming just Poisson noise and a circular sky aperture with inner and outer radius of 10 and 20 arc-seconds. Both methods produced similar SNR values, and we chose to adopt those from the LCOGT calculator. These SNR values shown in Table \ref{tabsnr} are strong functions of the phase of the moon. We incorporate this effect into the synthetic light curves by modulating the error bars with a lunar cycle. Gaussian noise is added to the data in the usual way,

\begin{equation}
\label{eqaddnoise}
F_{\nu} \left( \lambda , t \right) = F_{0 \nu} \left( \lambda , t \right) G \left( 1, \frac{1}{\mathrm{SNR}} \right),
\end{equation}

\noindent where the SNR value chosen is appropriate for the lunar phase in each week long segment, $F_{0\nu}$ are the perfect data. 

\begin{figure*}
\begin{center}
\includegraphics[scale=0.4,angle=0,trim=1.0cm 4.0cm 5cm 0cm,width=170mm,height=150mm]{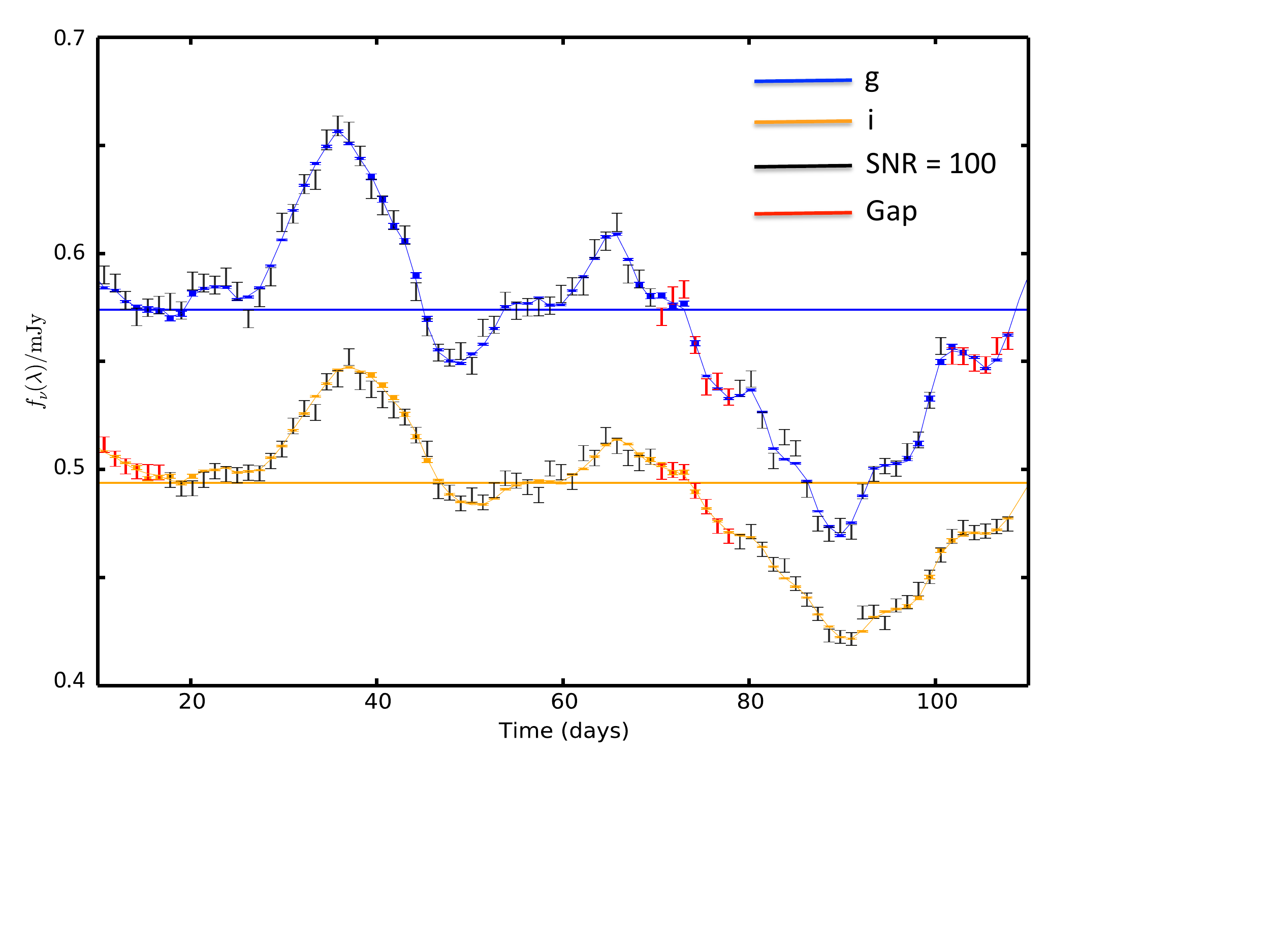}
\caption{The synthetic light curves used to test CREAM. Noise-free model light curves at g and i wavelengths are shown by the wavy blue and orange lines. The stars indicate a high quality light curve with mean SNR = 713 and 690 respectively but these vary depending on the lunar phase. We also test CREAM with SNR = 100 light curves in g and i shown by the black error bars. Red error bars are omitted from the SNR = 100 light curves to simulate week long data gaps as discussed in Section \ref{secgap}. The analysis of the high quality light curves, the SNR=100 light curve and the SNR = 100 light curves omitting the red points is discussed in Section \ref{secresults}.}
\label{figfake}
\end{center}
\end{figure*}

The final light curves in Figure \ref{figfake} show significant variability as expected from the observations of \citet{ma10}. Delays of the i with respect to the g light curve are clearly present and best seen by eye when comparing positions of individual peaks and troughs in the blue and orange lines.

\begin{table}
\center
\caption{Mean and RMS of the synthetic urgiz light curves. The RMS is a function of the DRW time scale $\tau_B$ and $\mathrm{SF}_\infty$ parameters measured by \citet{ma10}. These are a function of $z, M, \dot{M}$ and $\lambda$ and we quote the relevant values below.}
\begin{tabular}{ccccc}
\hline
Filter  & $\tau_\mathrm{B}$ & $SF_\infty$ & $\langle \mathrm{AB} \rangle $&  RMS \\
 &  (days) &  (mag) &  (mag) & (mag) \\
\hline 
u $3540 \AA$& 143 & 0.138 & 16.89 & 0.098 \\
g $4770 \AA$& 151 & 0.121 & 17.00 & 0.085 \\ 
r $6215 \AA$& 159 & 0.107 & 17.10 & 0.076 \\
i $7545 \AA$& 165 & 0.098 & 17.17 & 0.069 \\ 
z $8700 \AA$& 170 & 0.092 & 17.22 & 0.065 \\

\hline
\end{tabular}
\label{tabsynthprop}
\end{table}

\begin{table}
\center
\caption{The SNR's obtained from the LCOGT exposure time calculator for 1000 second observations on the ugriz filters of the 2m spectral cameras at airmass 1.3. The SNR is evaluated for 3 lunar phases. We show also the exposure time required to achieve SNR = 100. }
\begin{tabular}{ccccccc}
\hline
Filter  & \multicolumn{3}{c}{SNR for $t_\mathrm{exp} = 10^3 s$} & \multicolumn{3}{c}{$t_\mathrm{exp}$ for $\mathrm{SNR} = 100$} \\
& Dark & Half & Full & Dark & Half & Full\\
\hline 
u & 471 & 435 & 234 & 160 & 174 & 447 \\
g & 999 & 806 & 333 & 57  & 64  & 189 \\  
r & 967 &  889 & 443& 42  & 46  & 142 \\
i & 930 & 776 & 364 & 59  & 68  & 189 \\
z & 674 & 569 & 257 & 139 & 173 & 719 \\ 
\hline
\end{tabular}
\label{tabsnr}
\end{table}

\section{CREAM fitting results}
\label{secresults}

CREAM is applied to the g and i light curves plotted in Figure \ref{figfake}. We launch 3 independent chains with random initial values of $i$ between 0 and 90 degrees and $\mmdot$ between $10^7$ and $10^9 M_{\odot}^2 \mathrm{yr}^{-1}$. The Fourier $S_k$ and $C_K$ amplitudes start at 0. The offset and scaling terms for the echo light curve ($F_{\nu}\left( \lambda \right)$ and $\Delta F_{\nu} \left( \lambda \right)$) are started at the mean and RMS of each light curve. CREAM then cycles through the $2 \left( N_k + N_\lambda + 2 \right) + 1$ parameters and explores the parameter space subject to the priors discussed in Section \ref{secpriors}, stopping after $10^5$ iterations.

\begin{figure}
\center
\includegraphics[scale=0.55]{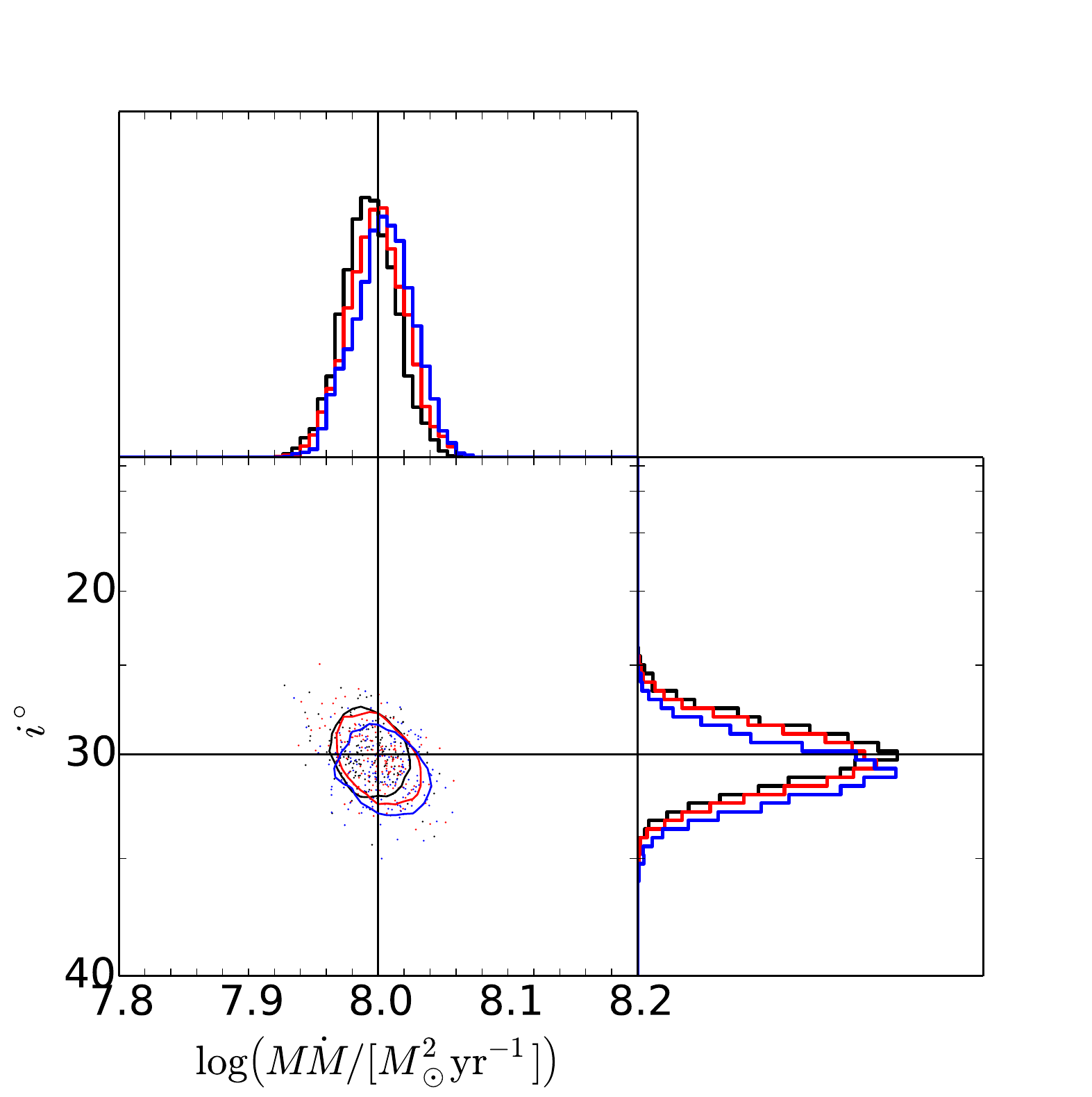}
\caption{Posterior probability distributions of $M\dot{M}$ and $i$ corresponding to the high SNR g and i light curves. The 3 colours indicate separate chains. Contours show the 1 $\sigma$ error regions. Black vertical and horizontal lines show the true parameters ($i = 30^\circ$ and $\mmdot = 10^8 M_{\odot}^2 \mathrm{yr}^{-1}$). }
\label{figcorplot_highsnr}
\end{figure}

CREAM fits to the various synthetic light curves are shown in Figures \ref{figresults_highsnr} to \ref{figresultsgap}. The inferred driving light curve $\Delta F_x \left( t \right)$ is shown in Panel a  and is convolved with the response functions $\psi \left( \tau | \lambda \right)$ in Panels b and d to yield the model g and i light curves respectively. The response functions are parameterised by $\mmdot$ and inclination. Using inclination and $\mmdot$, CREAM infers a delay distribution function from which we calculate the mean delays for the g and i light curves. Mean delays are proportional to $\left( \mmdot \right)^{1/3} \lambda^{4/3}$ and are annotated with uncertainties in Figures \ref{figresults_highsnr} to \ref{figresultsgap}. Mean delays are $2.62 \pm 0.05$ days and $4.78 \pm 0.08$ days respectively for the g and i light curves. It can be seen that CREAM achieves a superb fit to the data with tightly constrained response functions. The posterior probability distributions for the fits in Figures \ref{figresults_highsnr} to \ref{figresultsgap} are shown in Figures \ref{figcorplot_highsnr} to \ref{figcorplot_gap}. We also see from the red line in Panel a of Figures  \ref{figresults_highsnr} to \ref{figresultsgap} that CREAM is able to correctly infer the shape of the driving light curve with no data present. We see that the inferred model begins to deviate from the true driving light curve at times earlier than around 10 - 15 days before the first data point in g and i. The response functions for g and i (Panels b and d in Figure \ref{figresults_highsnr} are approximately 0 after a `look back' time of 15 days. The model at the time of the first data point is therefore insensitive to the driving light curve behaviour earlier than 15 days before the start of the observations.

Figure \ref{figcorplot_highsnr} shows the posterior distribution of the parameters $\mmdot$ and $i$, corresponding to the fit in Figure \ref{figresults_highsnr}. Colours represent individual chains. We note from the figure and the results in Table \ref{tabres} that CREAM is able to estimate inclination to $\pm 5^\circ$ and $\mmdot$ to $\pm 0.04$ dex. We regard this as a benchmark best case scenario and now explore the extent to which these results degrade with noisier observations.

\subsection{Less Favourable Observations}
\label{secsnr100}

\begin{figure}
\center
\includegraphics[scale=0.55]{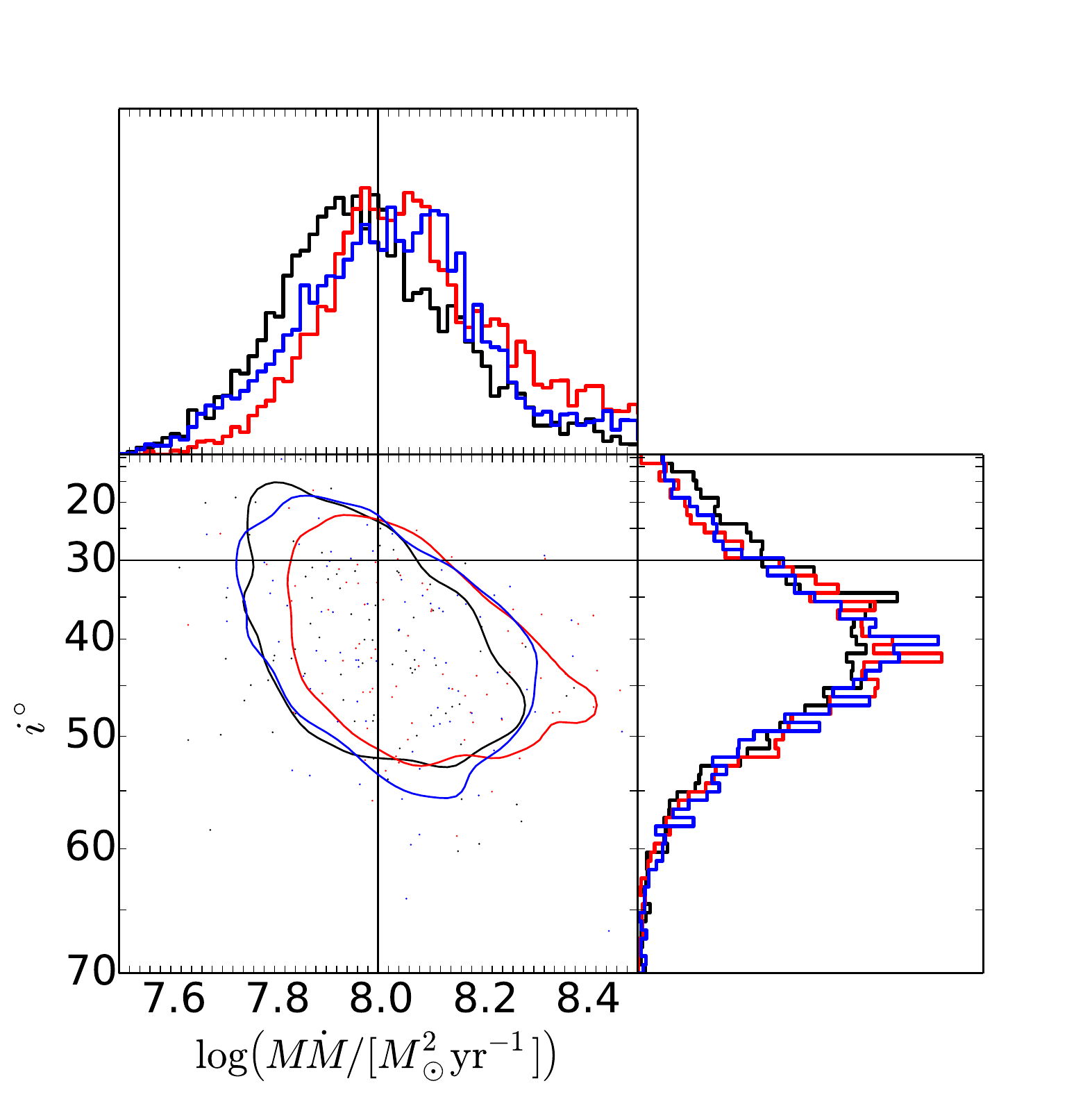}
\caption{As with Figure \ref{figcorplot_highsnr} but for the SNR = 100 g and i light curves.}
\label{figcorplot_snr100}
\end{figure}

Because systematic errors often limit CCD photometry, we consider CREAM's ability to recover $\mmdot$ and inclination from light curves with a poorer SNR. We use the same driving light curve as with the previous test and create the g and i response light curves using the prescription described in Section \ref{secsynthdat}. For this new set of g and i light curves, we generate error bars and perturb the data points assuming an SNR of 100. We show in Table \ref{tabsnr} the exposure time needed to obtain this SNR for our synthetic target. The CREAM fit to the noisier light curve is shown in Figure \ref{figresults_snr100} and posterior probability distributions in Figure \ref{figcorplot_snr100}. The uncertainty in inclination increases to $\pm 15^\circ$ and the uncertainty in $\mmdot$ rises to $\pm 0.19$ dex. Mean delays now exhibit uncertainties of 0.39 and 0.60 days for g and i respectively.

\begin{figure}
\center
\includegraphics[scale=0.45]{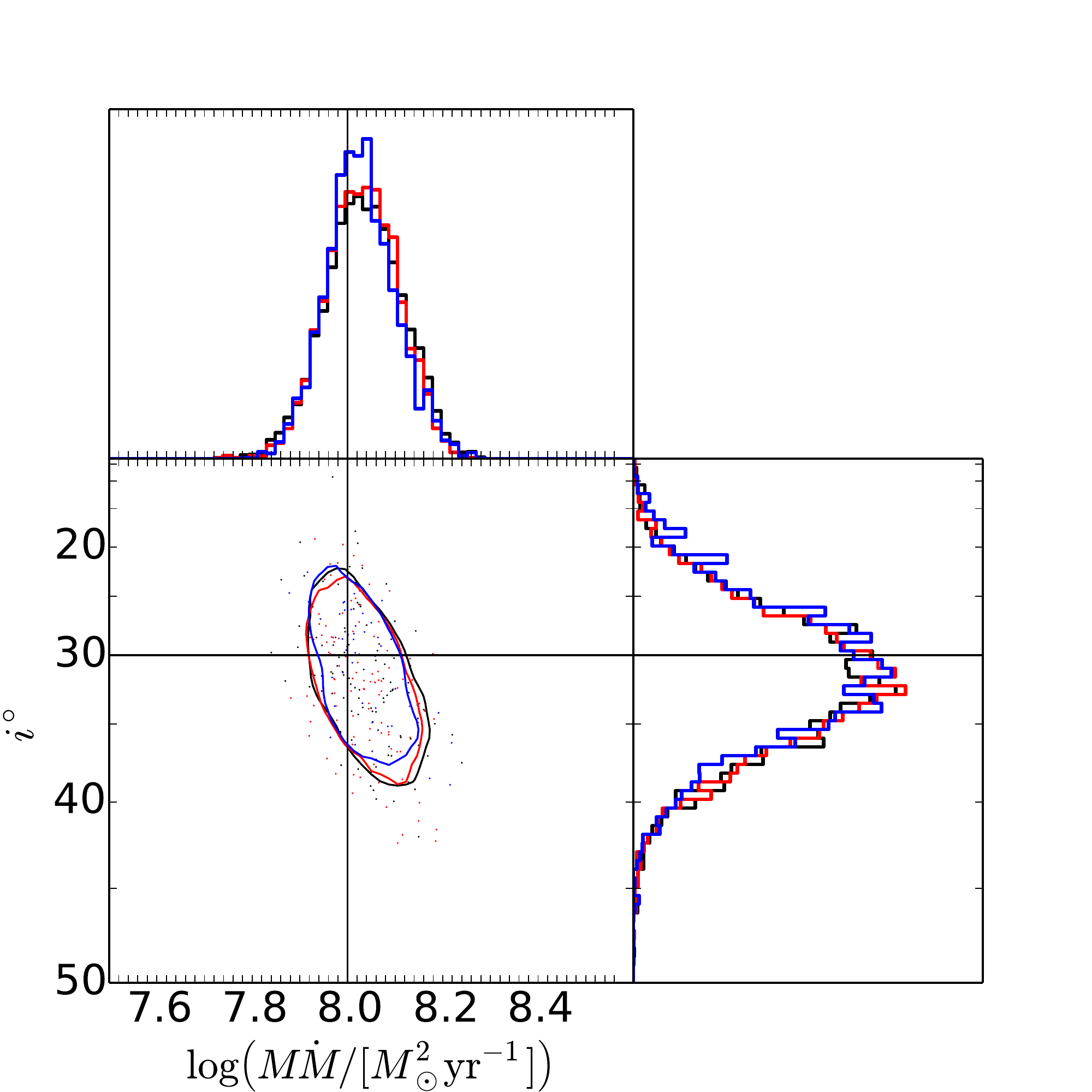}
\caption{As with Figure \ref{figcorplot_highsnr} but for CREAM fits to u,g,r,i and z light curves with SNR = 100 (Figure \ref{figresults_ugriz}). These light curves are driven by the same driving light curve as the g and i light curves in Figure \ref{figfake}.}
\label{figcorplot_ugriz}
\end{figure}

These results, while still accurate enough to be of scientific interest, are less attractive than those from the analysis of the higher quality light curves presented in Figure \ref{figresults_highsnr}. We therefore investigate how the results improve with the addition of SNR = 100 light curves in u, r and z. These are generated using the same procedure described in Section \ref{secsynthdat} and the relevant exposure times, and DRW parameters are given in Tables \ref{tabsnr} and \ref{tabsynthprop}. The same driving light curve is used to generate the additional u, r and z light curves as the earlier g and i data in Figure \ref{figfake}.  A fit to these light curves is shown in Figure \ref{figresults_ugriz}. Posterior probability distributions in Figure \ref{figcorplot_ugriz} show that adding the r,i and z light curves reduces the error by a factor of 2. We see that, for ugriz observations with SNR = 100, CREAM obtains an inclination accurate to $8.5^\circ$ and $\mmdot$ accurate to 0.1 dex.

We note a slight bias to high inclinations in Figure \ref{figcorplot_snr100}. We test this further by generating an additional 3 driving light curves and obtaining g and i light curves for each using the prescription in Section \ref{secsynthdat}. The prosterior probabilities for $\mmdot$ and inclination are shown for each of these drivers in Figure \ref{figcorplot_multidrive}. We see that the inferred inclination is sometimes above and sometimes below the true value of $30^\circ$. This scatter is expected as the 2 parameter $1 \sigma$ contours should include the true model in 68 \% of the driving light curves tested.

\subsection{Data Gaps}
\label{secgap}

\begin{figure}
\center
\includegraphics[scale=0.55]{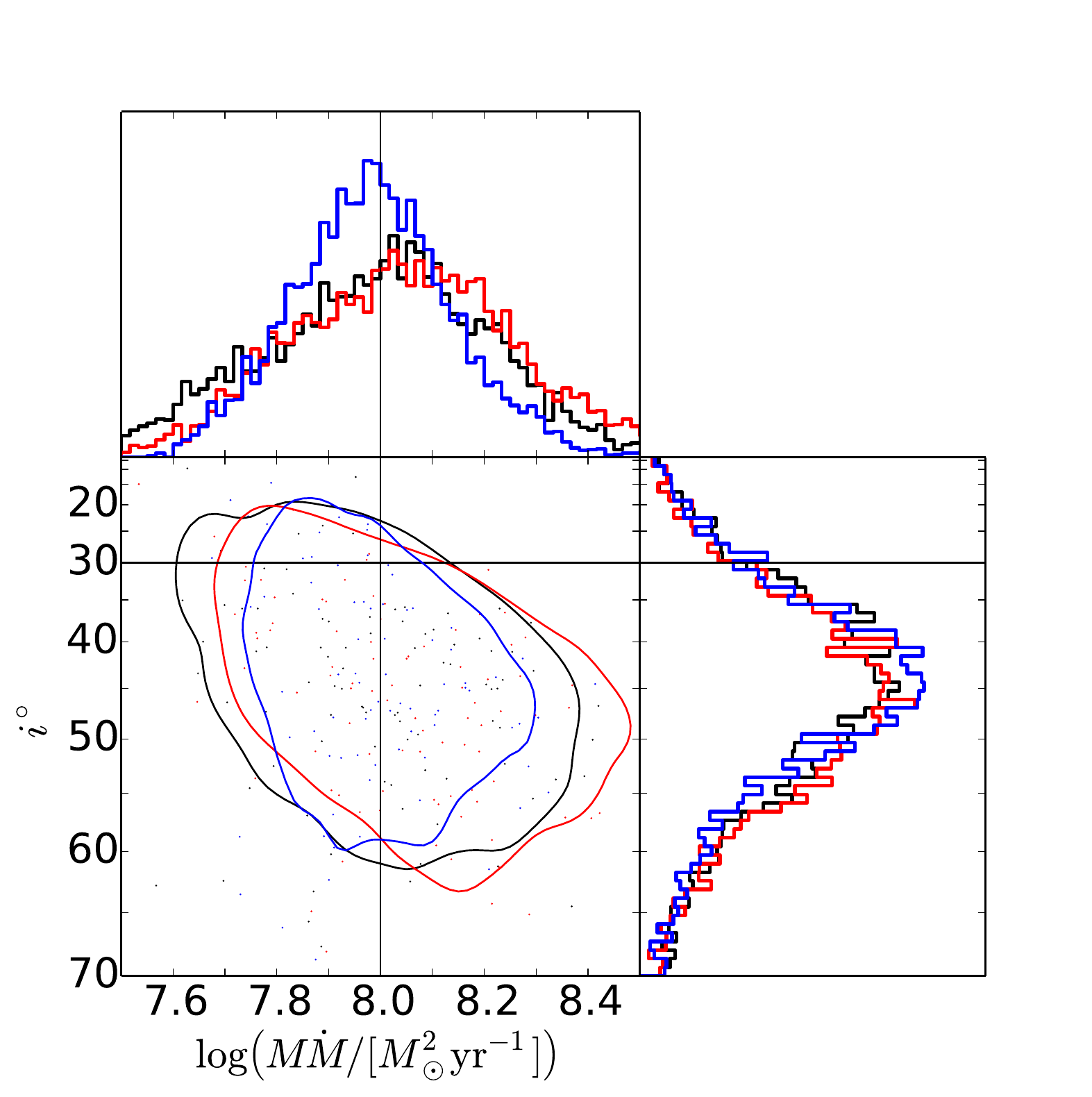}
\caption{As with Figure \ref{figcorplot_highsnr} but for the SNR = 100 g and i light curves including the 4 week-long data gaps (Figure \ref{figresultsgap}.}
\label{figcorplot_gap}
\end{figure}

\begin{figure}
\center
\includegraphics[scale=0.45]{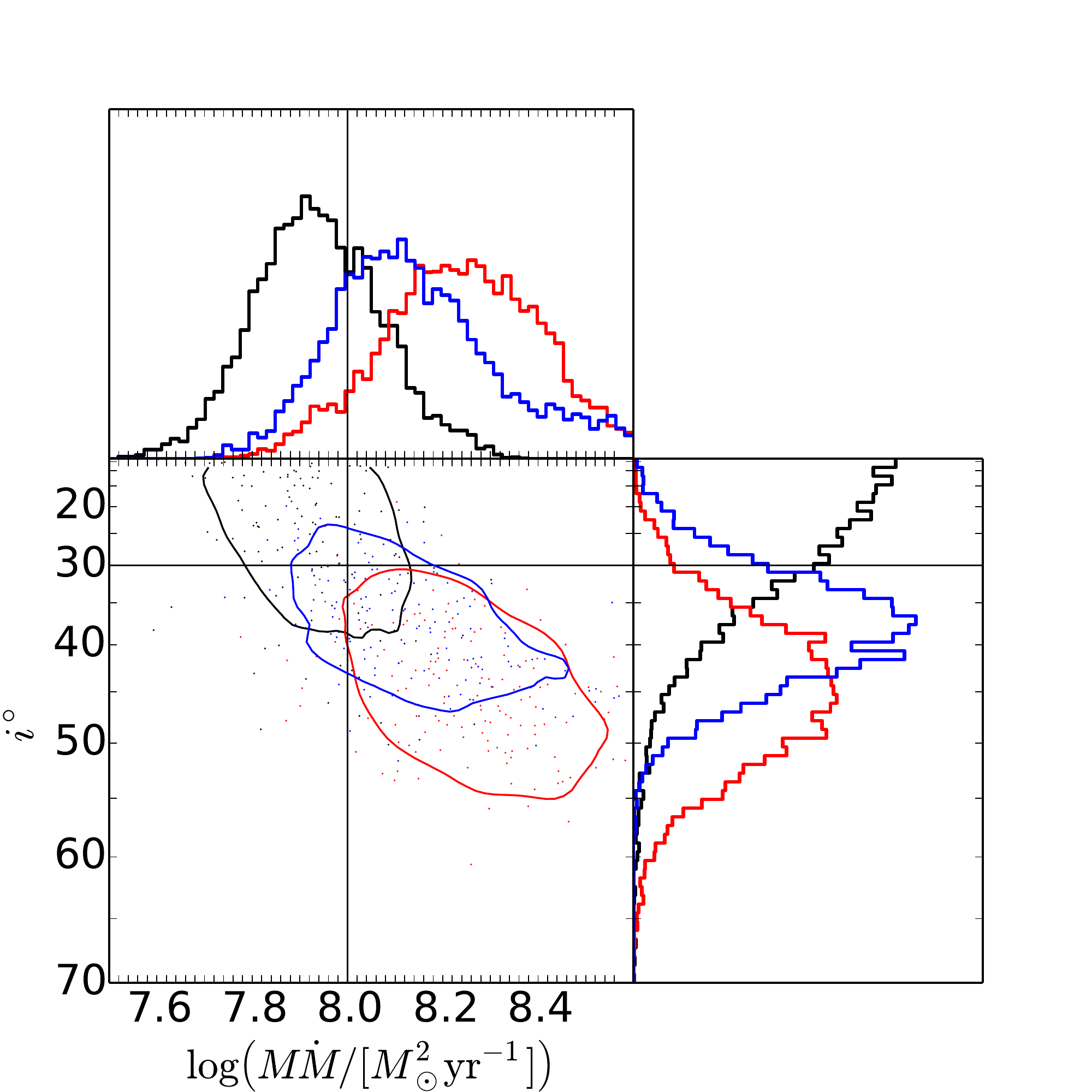}
\caption{As with Figure \ref{figcorplot_highsnr} but for CREAM fits to g and i light curves with colours now showing 3 different driving light curves. }
\label{figcorplot_multidrive}
\end{figure}

The light curves in Figure \ref{figfake}, while not equally sampled, are fairly regular with mean 1 day cadence. To investigate the possible impact of  data gaps (e.g due to equipment failure or episodes of poor weather), we remove 4 one-week-long segments from the light curves, the first week from the start of the i light curve, the second and third weeks from both g and i light curves simultaneously at 50 - 57 days, and the final week from the end of the i light curve. We perform this test on the noisier g and i light curves with SNR = 100. This is the worst case observing scenario we consider. The CREAM fit to light curves with the data gaps are shown in Figure \ref{figresultsgap}. The posterior probability distributions for the inclination and $\mmdot$ are plotted in Figure \ref{figcorplot_gap}. The error envelope for the inferred driving and echo light curves expands inside the gaps as appropriate. It is apparent that CREAM's ability to recover inclination is adversely affected when the gaps are included in the light curves. The uncertainty in the inclination now rises to $\pm 18^\circ$ and the uncertainty in $\mmdot$ rises to 0.24 dex. Light curve mean delay uncertainties for g and i again increase to 0.74 and 0.95 days respectively.

\begin{figure}
\center
\includegraphics[scale=0.65]{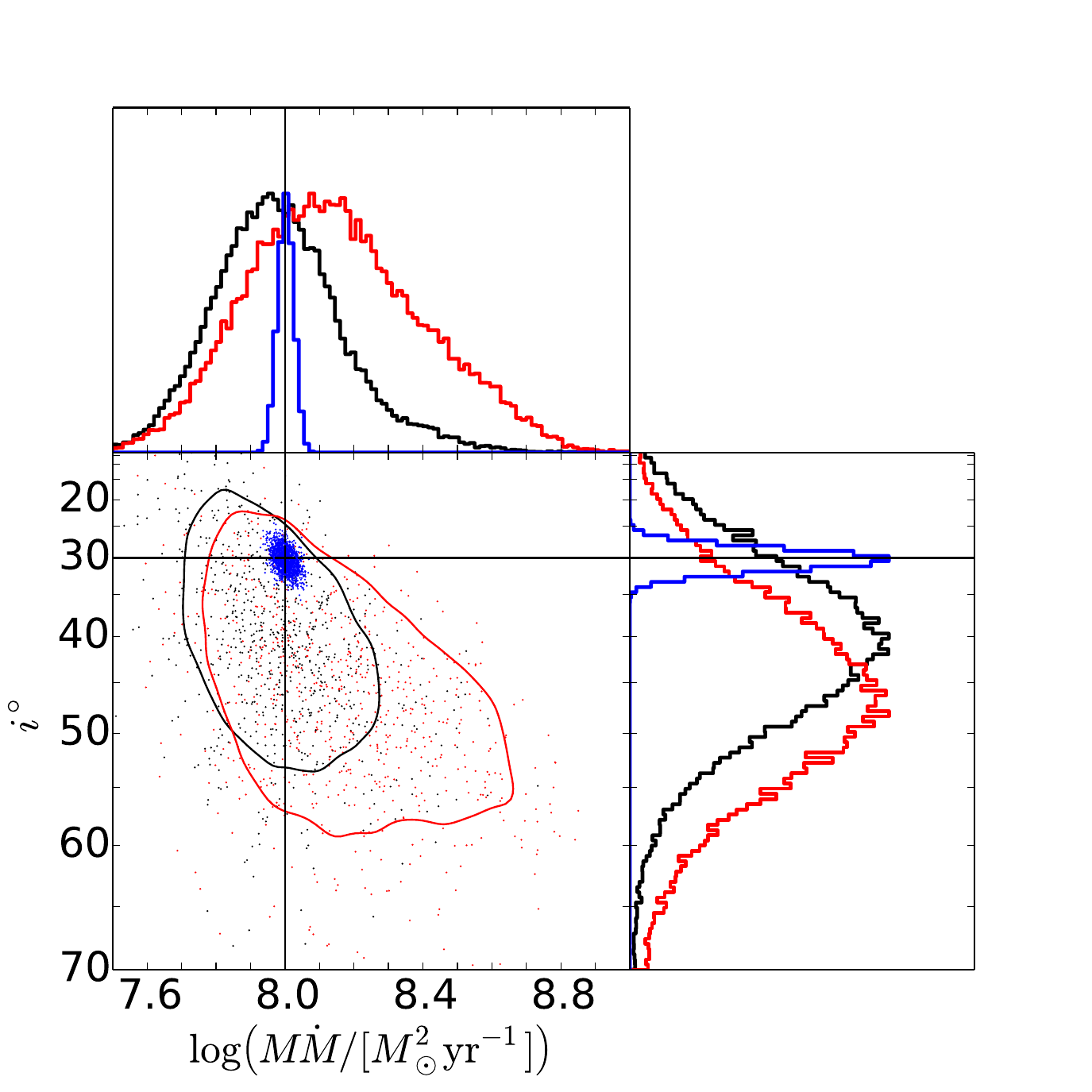}
\caption{Here we add together the results from all 3 chains in Figures \ref{figcorplot_highsnr}, \ref{figcorplot_snr100}, \ref{figcorplot_gap}. Blue represents the high SNR light curves, black is the SNR = 100 sample with no gaps and red includes the four week-long gaps in the light curve.}
\label{figcorplot_comb}
\end{figure}

\begin{figure*}
\center
\begin{center}
\includegraphics[scale=1.3,angle=0,trim=0cm 0cm 0cm 0cm]{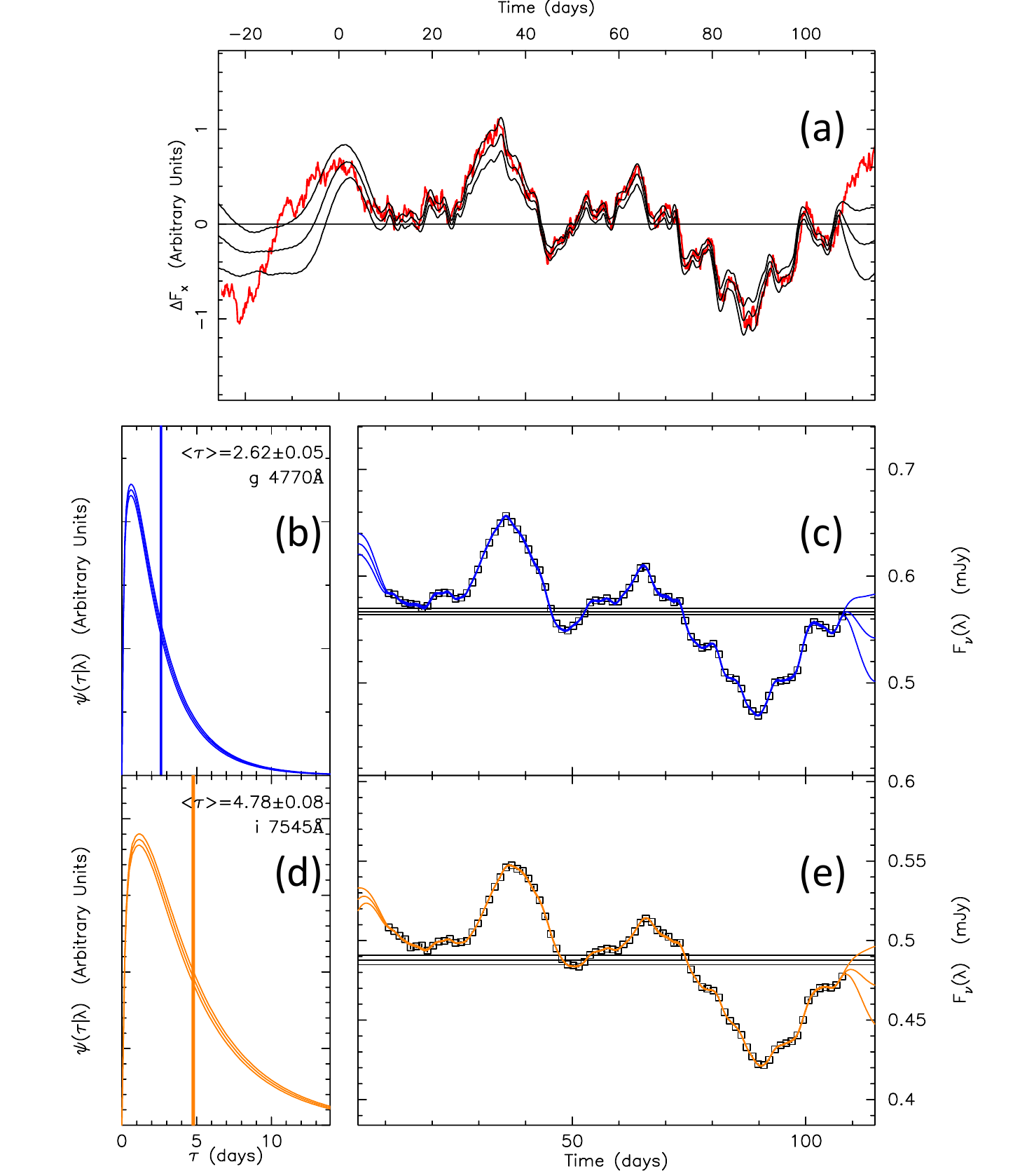}
\caption{The top plot shows the reconstructed driving light curve (Equation \ref{eqdrive}). The plots below this show (left) the inferred response function (Figure \ref{figtf}) with vertical lies showing the mean time delay $\langle \tau \rangle$. The right plots show the response light curves in the g and i filters including 1 $\sigma$ uncertainty envelopes. Horizontal lines indicate the offset parameters and uncertainty envelopes for the response light curves. The red line in Panel a shows the driving light curve used to generate the synthetic g and i light curves, normalised to the rms of the inferred driving light curve.}
\label{figresults_highsnr}
\end{center}
\end{figure*}

\begin{figure*}
\center
\begin{center}
\includegraphics[scale=1.3,angle=0,trim=0cm 0cm 0cm 0cm]{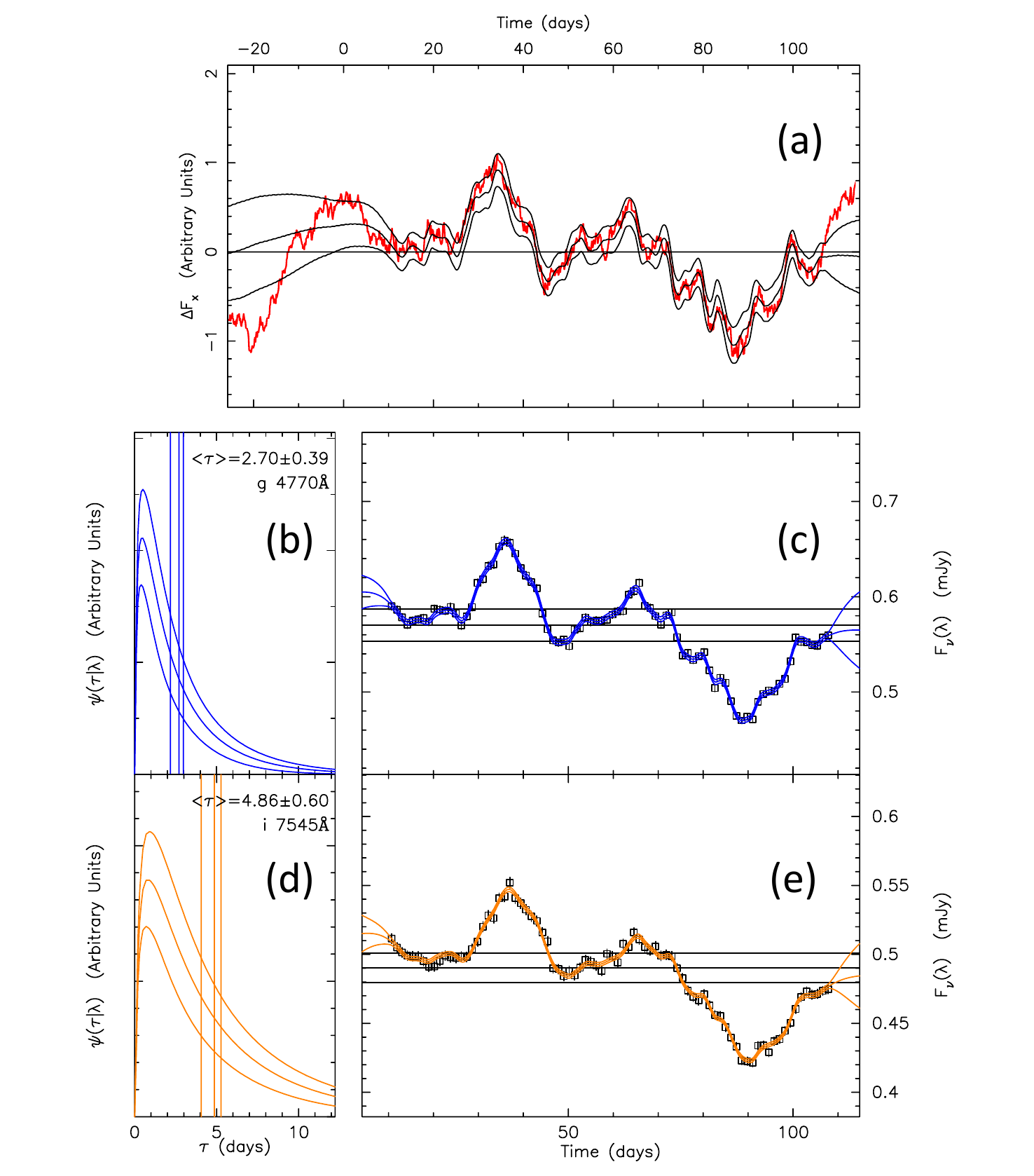}
\caption{As with Figure \ref{figresults_highsnr} but showing CREAM applied to the noiser SNR = 100 light curve data.}
\label{figresults_snr100}
\end{center}
\end{figure*}

\begin{figure*}
\center
\begin{center}
\includegraphics[scale=1.3,angle=0,trim=0cm 0cm 0cm 0cm]{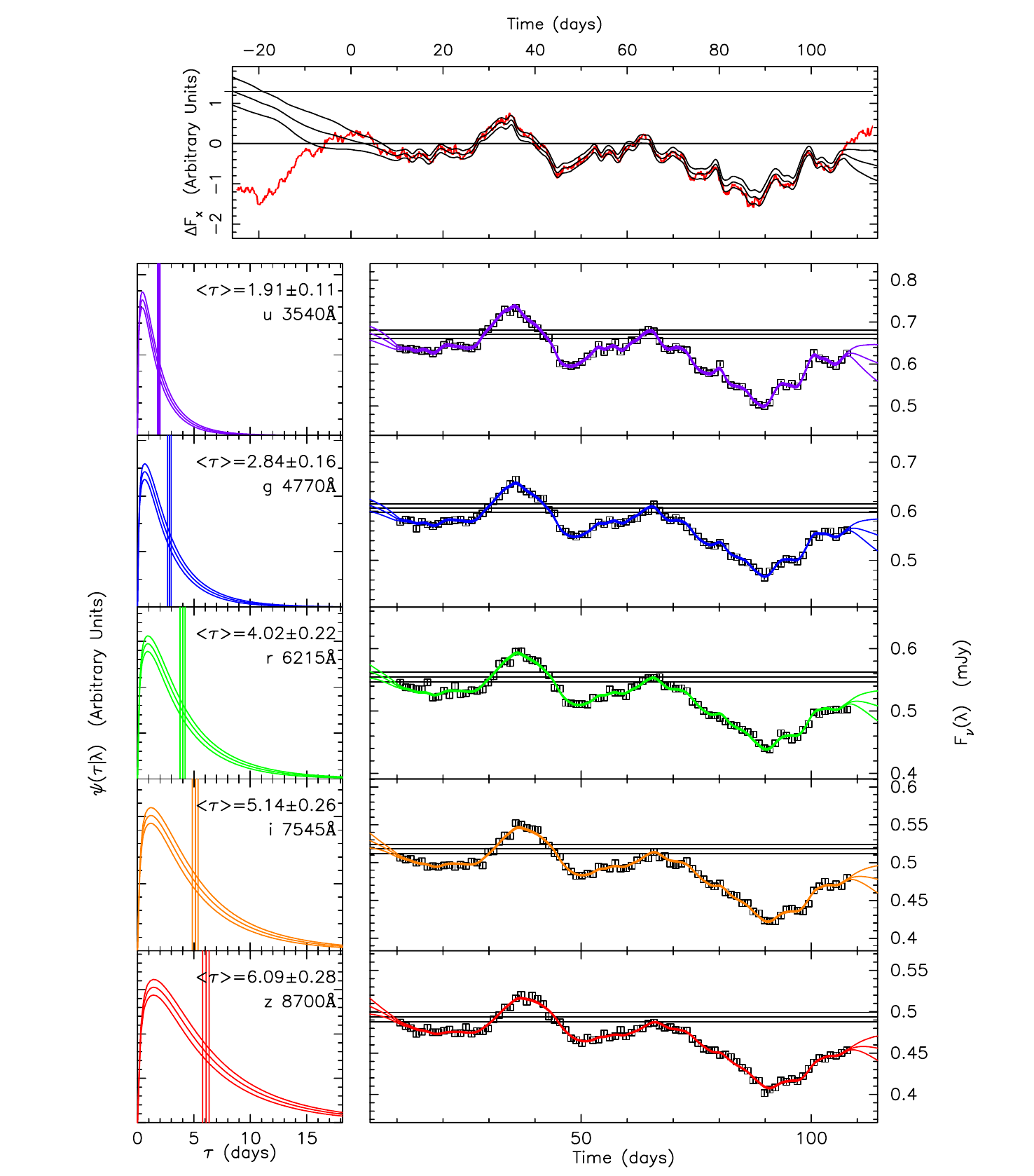}
\caption{As with Figure \ref{figresults_highsnr} but showing CREAM applied to the SNR = 100 light curves in ugriz.}
\label{figresults_ugriz}
\end{center}
\end{figure*}

\begin{figure*}
\center
\begin{center}
\includegraphics[scale=1.3,angle=0,trim=0cm 0cm 0cm 0cm]{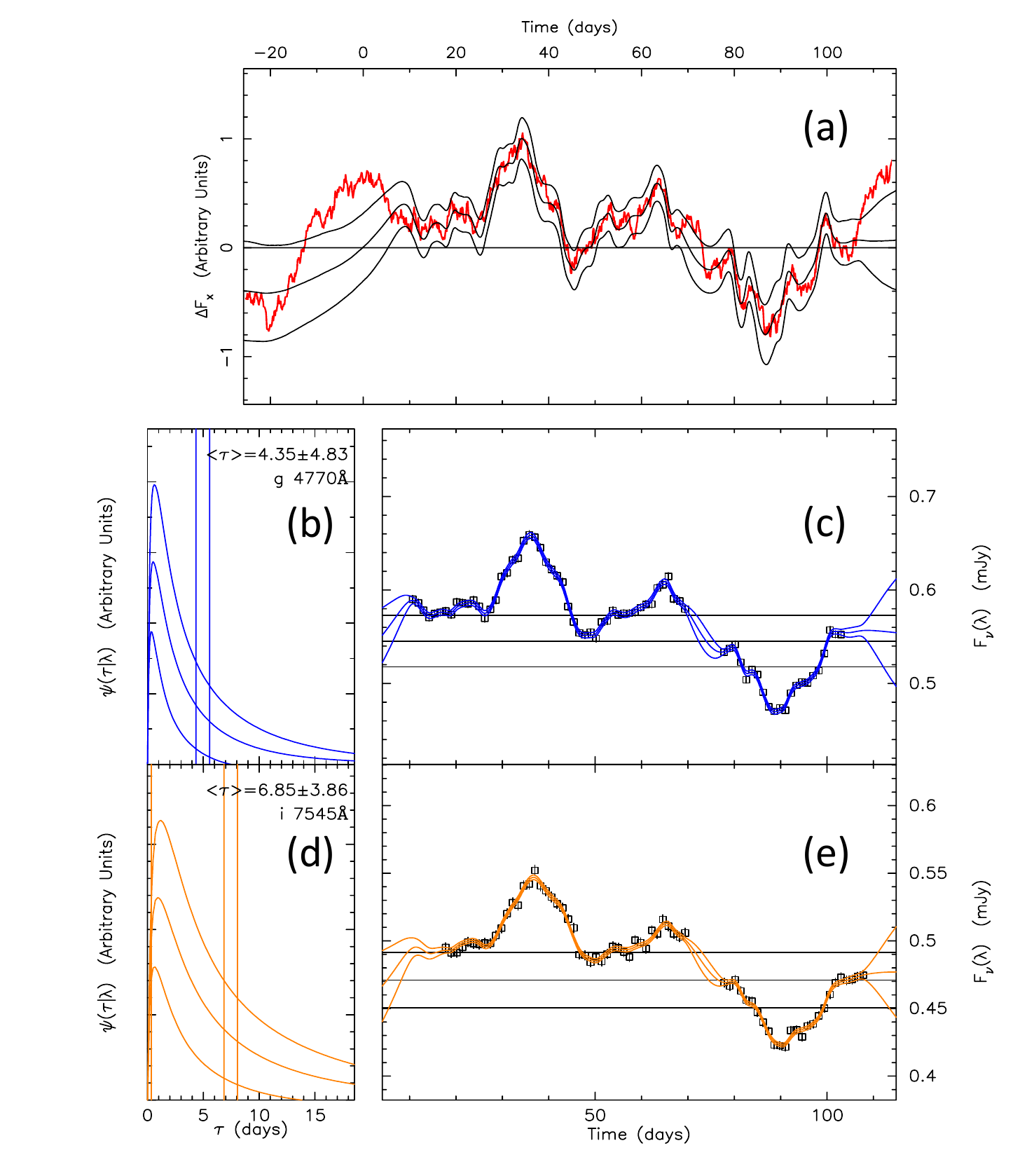}
\caption{As with Figure \ref{figresults_highsnr} but including the four - 1 week gaps in the light curves. Gaps occur at the start of the g light curve, in g and i at 50 - 57 days and at the end of the g light curve.}
\label{figresultsgap}
\end{center}
\end{figure*}

A summary plot showing the posterior probabilities for inclination and $\mmdot$ is displayed for ease of comparison between the various tested observing strategies in Figure \ref{figcorplot_comb}. The accuracy achieved in all these tests would be of scientific interest if achieved in practice.

\subsection{High Frequency Resolution Limits: $\omega_\mathrm{hi}$}

CREAM fits consume computer time due to the large number of parameters in the Fourier series (Equation \ref{eqdrive}). We ran tests lowering $\omega_\mathrm{hi}$ to speed up the fitting. We find no effect on the $\cos i$ or $\log \mmdot$ parameter estimates so long as the driving light curve is evaluated with upper frequencies of 0.5 cycle / day or higher. This approximately matches the inferred $\langle \tau \rangle$ for the i light curve in Figure 7 suggesting future observations at shorter wavelengths than those presented here may need to be run with larger values of $\omega_\mathrm{hi}$ higher resolution.

\FloatBarrier

\FloatBarrier

\begin{table}
\begin{center}
\caption{The $M\dot{M}$ and inclination estimates for each of the 3 CREAM runs to the light curves shown in the figures stated. A fit to the light curves with high, lunar-phase-dependant SNR's is shown in Figure \ref{figresults_highsnr}. The g and i SNR = 100 light curves are fitted in Figure \ref{figresults_snr100}. The ugriz light curves are fitted in Figure \ref{figresults_ugriz}. CREAM is fitted to g and i light curves from 3 different drivers and the posterior probabilities for $\mmdot$ and inclination are shown in Figure \ref{figcorplot_multidrive}. CREAM runs on the SNR = 100 light curves with the 4 week-long gaps are displayed in Figures \ref{figresultsgap} and \ref{figcorplot_gap}.}
\begin{tabular}{ccc}
\hline
Light Curve Set  & $\log \left(\mmdot \right)$  & $i$ (deg) \\
\hline
Figures \ref{figresults_highsnr} and \ref{figcorplot_highsnr} & 8.00 $\pm$ 0.03 & 29.28 $\pm$ 5.04   \\
  & 8.00 $\pm$ 0.04 & 30.84 $\pm$ 4.90   \\
  & 8.01 $\pm$ 0.03 & 31.45 $\pm$ 4.52   \\
Figures \ref{figresults_snr100} and \ref{figcorplot_snr100} & 7.97 $\pm$ 0.17 & 41.58 $\pm$ 13.32  \\
  & 8.06 $\pm$ 0.19 & 42.40 $\pm$ 14.12  \\
  & 8.03 $\pm$ 0.18 & 43.07 $\pm$ 14.62  \\
Figures \ref{figresults_ugriz} and \ref{figcorplot_ugriz} & 8.03 $\pm$ 0.08 & 31.9 $\pm$ 8.48  \\
  & 8.02 $\pm$ 0.08 & 31.97 $\pm$ 7.84  \\
  & 8.03 $\pm$ 0.10 & 31.31 $\pm$ 7.96  \\   
Figure \ref{figcorplot_multidrive} & 8.15 $\pm$ 0.18 & 38.30 $\pm$ 8.52  \\
  & 8.26 $\pm$ 0.18 & 45.10 $\pm$ 9.49  \\
  & 7.93 $\pm$ 0.13 & 18.77 $\pm$ 15.62  \\  
Figures \ref{figresultsgap} and \ref{figcorplot_gap} & 8.01 $\pm$ 0.24 & 45.99 $\pm$ 18.02  \\
  & 7.98 $\pm$ 0.15 & 47.78 $\pm$ 16.33  \\
  & 8.05 $\pm$ 0.23 & 46.61 $\pm$ 17.23  \\
\hline
\end{tabular}
\label{tabres}
\end{center}
\end{table}

\section{Discussion and Conclusions}
\label{secdiscussion}

We have introduced a new approach (CREAM) to measure the continuum lag distribution of reverberating AGN accretion discs. The model assumes a flat optically-thick accretion disc that undergoes temperature perturbations due to irradiation by a variable driving point source (lamp post) located above the disc centre. Time lags increase with $\lambda$ and arise due to light travel time effects that are longer for the cooler, longer-wavelength-emitting regions located at larger radii. Observations of the power spectrum of X-ray light curves suggest that X-rays may lead the variability \citep{mc14} making them a strong candidate for the lamp post. This variability exhibits a random walk power spectrum $P(f) \propto f^{-2}$ that we adopt to constrain our model parameters. 

This approach is similar to SPEAR and JAVELIN \citep{zu11} which use DRW light curves in an attempt to model an input light curve (usually the shortest wavelength light curve is a proxy for the driving light curve) and model each echo light curve individually assuming a top hat delay distribution. CREAM can infer both $\mmdot$ and disc inclination parameters of the delay distribution, and the driving light curve with no input driver required.

To test CREAM we generated synthetic optical light curves using the thermal reprocessing model assuming a delay distribution appropriate for an $\mmdot = \mathrm{10^8 M_{\odot} ^2 yr^{-1}}$ accretion disc inclined at $\mathrm{30^{\circ}}$. The light curves are randomly sampled with mean 1 day cadence over 100 days and mimic the SNR expected from 1000s observations in g and i with a 2m telescope. 

We emphasise that CREAM does not require data to act as a proxy for the variable irradiation (the driving light curve). Since we do not fully understand the origin of continuum variability \citep{lo14} this is an exciting feature of the code. We anticipate comparing CREAM's inference of the driving light curve with various UV and X-ray light curves to investigate which of these bands CREAM best agrees with. 

In this work, we use tests with simulated data to demonstrate CREAM's ability to recover $M\dot{M}$ to $\pm 0.04$ dex and disc inclinations to $\pm 5^\circ$ with observations at 1 day cadence in g and i filters. We note also that with noisier light curves (SNR = 100) our inferred inclination uncertainty increases to $\pm 15^\circ$. The uncertainty in $\mmdot$ increases to $0.2$ dex. If we include overlapping SNR=100 light curves in u,r and z we are able to infer $\mmdot$ to $\pm 0.1$ dex and inclination to $\pm 8.5^\circ$. Any of these would be of scientific interest.

While CREAM recovers delay distributions for continuum - continuum lags, it does so by applying a number of simplifications. CREAM assumes that variability in continuum light curves arises solely due to lamp post thermal reprocessing. There may in fact be contaminating contributions to light curve variability from the BLR. \citet{ko01} note the presence of diffuse continuum emission from the BLR. The diffuse-cloud-continuum emission is a combination of reflected accretion disc continuum and thermal-non-line emission. If present, diffuse continuum emission might increase the time delays, causing CREAM to overestimate the $\mmdot$ and mean delay. This would also affect the mean delays as obtained by SPEAR and CCF. 

In an upcoming paper, we will apply CREAM to 19 overlapping, multiwavelength continuum light curves taken by the STORM collaboration \citep{ed15,ro15}. We aim to both measure the disc inclination, accretion rate and infer the shape of the driving light curve; testing the agreement between the X-ray light curve and CREAM's inferred driving light curve to determine how the X-rays are related to continuum variability.

In summary, the ability to constrain accretion disc delay distribution functions affords CREAM many potential applications. Not least of these is the information we can obtain on the accretion rate and disc inclination. Multi-wavelength observations of NGC 5548 \citep{ro15,ed15} offer real datasets on which to apply CREAM to measure the accretion disc parameters. Finally, since no data are needed to act as the driver of variability, the driving light curve can be recovered from CREAM fits to observed echo light curves and compared to observations at different wavelengths to help identify the true driver of AGN variability.

\section{Acknowledgements}
We thank the referee for the helpful comments and suggestion made during the submission of this manuscript.
D.A.S acknowledges the support of the Science and Technologies Funding Council studentship.
K.H acknowledges support from the UK Science and Technology Facilities Council (STFC) consolidated grant to St.Andrews (ST/M001296/1).

\bibliography{cream_final} 
\bibliographystyle{mn2e}

\end{document}